\documentclass{article}
\usepackage{ismir,amsmath,cite,url}
\usepackage{graphicx}
\usepackage{color}

 \usepackage{amssymb, latexsym}
 \usepackage{amsthm}
 \usepackage{bm}
 \usepackage[mathscr]{eucal}
 \usepackage{enumerate}

\title{Calibration of a two-state pitch-wise HMM method for note segmentation in Automatic Music Transcription systems}

\multauthor
{Dorian Cazau$^1$ \hspace{1cm} Yuancheng Wang$^2$ \hspace{1cm} Olivier Adam$^3$} { \bfseries{Qiao Wang$^2$ \hspace{1cm} Gr\'egory Nuel$^4$}\\
$^1$ Lab STIC, ENSTA-Bretagne\\
$^2$ Southeast China\\
$^3$  Institut d'Alembert, UPMC\\
$^4$  LPMA, UPMC\\
{\tt\small dorian.cazau@ensta-bretagne.fr}
}
\def\authorname{First author, Second author, Third author, Fourth author, Fifth author}

\usepackage{multirow}

\begin{document}

\maketitle
\begin{abstract}
Many methods for automatic music transcription involves a multi-pitch estimation method that estimates an activity score for each pitch. A second processing step, called note segmentation, has to be performed for each pitch in order to identify the time intervals when the notes are played. In this study, a pitch-wise two-state on/off first-order Hidden Markov Model (HMM) is developed for note segmentation. A complete parametrization of the HMM sigmoid function is proposed, based on its original regression formulation, including a parameter $\alpha$ of slope smoothing and $\beta$ of thresholding contrast. A comparative evaluation of different note segmentation strategies was performed, differentiated according to whether they use a fixed threshold, called ``Hard Thresholding" (HT), or a HMM-based thresholding method, called ``Soft Thresholding" (ST). This evaluation was done following MIREX standards and using the MAPS dataset. Also, different transcription scenarios and recording natures were tested using three units of the Degradation toolbox. Results show that note segmentation through a HMM soft thresholding with a data-based optimization of the $\{\alpha , \beta \}$ parameter couple significantly enhances transcription performance.
\end{abstract}

\section{Introduction}\label{sec:introduction}

Work on Automatic Music Transcription (AMT) dates back more than 30 years \cite{Moorer1977}, and has known numerous applications in the fields of music information retrieval, interactive computer systems, and automated musicological analysis \cite{Klapuri2004b}. Due to the difficulty in producing all the information required for a complete musical score, AMT is commonly defined as the computer-assisted process of analyzing an acoustic musical signal so as to write down the musical parameters of the sounds that occur in it, which are basically the pitch, onset time, and duration of each sound to be played. This task of ``low-level" transcription, to which we will restrict ourselves in this study, has interested more and more researchers from different fields (e.g. library science, musicology, machine learning, cognition), and has been a very competitive task in the MIR (Music Information Retrieval) community \cite{Mirex2011} since the early 2000s. Despite this large enthusiasm for AMT challenges, and several audio-to-MIDI converters available commercially, perfect polyphonic AMT systems are out of reach of today's technology. 

The diversity of music practice, as well as supports of recording and diffusion, makes the task of AMT very challenging indeed. These variability sources can be partitioned based on three broad classes: 1) instrument based, 2) music language model based and 3) technology based. The first class covers variability from tonal instrument timbre. All instruments possess a specific acoustic signature, that makes them recognizable among different instruments playing a same pitch. This timbre is defined by acoustic properties, both spectral and temporal, specific to each instrument. The second class includes variability from the different ways an instrument can be played, that vary with the musical genre (e.g. tonality, tuning, rhythm), the playing techniques (e.g. dynamics, plucking modes), and the personal interpretations of a same piece. These first two classes induce a high complexity of note spectra over time, whose non-stationary is determined both by the instrument and the musician playing characteristics. The third class includes variability from electromechanics (e.g. transmission channel, microphone), environment (e.g. background noise, room acoustics, distant microphone), data quality (e.g. sampling rate, recording quality, audio codec/compression). For example, in ethnomusicological research, extensive sound datasets currently exist, with generally poor quality recordings made on the field, while a growing need for automatic analysis appears \cite{Moelants2007,Lidy2010,Cornelis2010,Six2013}.

Concerning AMT methods, many studies have used rank reduction and source separation methods, exploiting both the additive and oscillatory properties of audio signals. Among them, spectrogram factorization methods have become very popular, from the original Non-negative Matrix Factorization (NMF) to the recent developments of the Probabilistic Latent Component Analysis (PLCA) \cite{Benetos2013,Arora2014}. PLCA is a powerful method for Multi-Pitch Estimation (MPE), representing the spectra as a linear combination of vectors from a dictionary. Such models take advantage of the inherent low-rank nature of magnitude spectrograms to provide compact and informative descriptions. Their output generally takes the form of a pianoroll-like matrix showing the ``activity" of each spectral basis against time, that is itself discretized into successive time frame of analysis (of the order of magnitude of 11 ms). From this activity matrix, the next processing step in view of AMT is note segmentation, that aims to identify for each pitch the time intervals when the notes are played. To perform this operation, most spectrogram factorization-based transcription methods \cite{Grindlay2011,Mysore2009,Dessein2010} use a simple threshold-based detection of the note activations from the pitch activity matrix, followed by a minimum duration pruning. One of the main drawback of this PLCA method with a simple threshold is that all successive frame are processed independently from one another, and thus temporal correlation between successive frames is not modeled. One solution that has been proposed is to jointly learn spectral dictionaries as well as a Markov chain that describes the structure of changes between these dictionaries  \cite{Mysore2009,Nakano2010,Benetos2013}. 

In this paper, we will focus on the note segmentation stage, using a pitch-wise two-state on/off first-order HMM, initially proposed by Poliner et al. \cite{Poliner2007} for AMT. This HMM allows taking into account the dependence of pitch activation across time frames. We review the formalism of \cite{Poliner2007}'s model, including a full parametrization of the sigmoid function used to map HMM observation probabilities into the $[0,1]$ interval, with a term $\alpha$ of slope smoothing and $\beta$ of thresholding contrast. After demonstrating the relevance of an optimal adjustment of these parameters for note segmentation, a supervised approach to estimate the sigmoid parameters from a learning corpus is proposed. Also, the degradation toolbox \cite{Mauch2013} was used to build three ``degraded" sound datasets that have allowed to evaluate transcription performance on real life types of audio recordings, such as radio broadcast and MP3 compressed audio, that are almost never dealt with in transcription studies.

\section{Methods}

\subsection{Background on PLCA}

PLCA is a probabilistic factorization method \cite{Smaragdis2006} based on the assumption that a suitably normalized magnitude spectrogram, $V$, can be modeled as a joint distribution over time and frequency, $P(f,t)$, with $f$ is the log-frequency index and $t=1,\ldots,T$ the time index with $T$ the number of time frames. This quantity can be factored into a frame probability $P(t)$, which can be computed directly from the observed data (i.e. energy spectrogram), and a conditional distribution over frequency bins $P(f|t)$, as follows \cite{Cazau2013}

\begin{equation}\label{EqPLCA}
	P(f|t)= \sum\limits_{p,m} P(f | p,m) P(m|p,t) P(p|t)
\end{equation}

where $P(f|p,m)$ are the spectral templates for pitch $p=1,\ldots,N_p$ (with $N_p$ the number of pitches) and playing mode $m$, $P(m|p,t)$ is the playing mode activation, and $P(p|t)$ is the pitch activation (i.e. the transcription). In this paper, the playing mode $m$ will refer to different playing dynamics (i.e. note loudness). To estimate the model parameters $P(m|p,t)$ and $P(p|t)$, since there is usually no closed-form solution for the maximization of the log-likelihood or the posterior distributions, iterative update rules based on the Expectation-Maximization (EM) algorithm \cite{Dempster1977} are employed (see \cite{Benetos2012c} for details). The pitch activity matrix $P(p,t)$ is deduced from $P(p|t)$ with the Bayes' rule

\begin{equation}
P(p,t) =  P(t)P(p|t) 
\end{equation}

PLCA note templates are learned with pre-recorded isolated notes, using a one component PLCA model (i.e. $m$ = 1 in Eq.~(\ref{EqPLCA}). Three different note templates per pitch are used during MPE. In this paper, we use the PLCA-based AMT system developed by Benetos and Weyde \cite{Benetos2015b}\footnote{Codes are available at \url{https://code.soundsoftware.ac.uk/projects/amt_mssiplca_fast}.}. 

In the following, for $p=1,\ldots,N_p$ and $t=1,\ldots,T$, we define the logarithmic pitch activity matrix as

\begin{equation}
X_{p,t} = \log \big( P(p,t) \big)
\end{equation}

\subsection{Note segmentation strategies}

\subsubsection{HT: Hard Thresholding}\label{PLCAtheory}

The note segmentation strategy HT consists of a simple thresholding $\beta_\text{HT}$ of the logarithmic pitch activity matrix $X(p,t)$, as it is most commonly done in spectrogram factorization-based transcription or pitch tracking systems, e.g. in \cite{Mysore2009,Dessein2010,Grindlay2011}. This HT is sometimes combined with a minimum duration constraint with typical post filtering like ``all runs of active pitch of length smaller than k are set to 0".

%

%
%

\subsubsection{ST: Soft Thresholding}\label{TwoStateHMMmodels}

In this note segmentation strategy, initially proposed by Poliner and Ellis \cite{Poliner2007}, each pitch $p$ is modelled as a two-state on/off HMM, i.e. with underlying states $q_t$ $\in$ $\{0,1\}$ that denote pitch activity/inactivity. The state dynamics, transition matrix, and state priors are estimated from our ``directly observed" state sequences, i.e. the training MIDI data, that are sampled at the precise times corresponding to the analysis frames of the activation matrix.

\begin{figure}
\centering
\def\svgwidth{7cm}
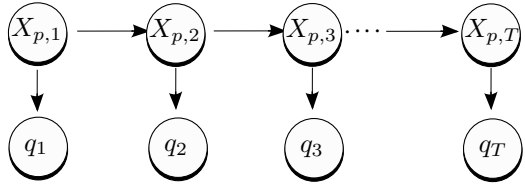
\caption{Graphical representation of the two-state on/off HMM. $q_t \in \{0,1\}$ are the underlying states label at time t, and $o_t$ the the probability observations.}
\label{DOHMM_StateTransition}
\end{figure}



For each pitch $p$, we consider an independent HMM with observations $X_{p,t}$, that are actually observed, and hidden binary Markov sequence $Q=q_1,\ldots,q_T$, illustrated in figure \ref{DOHMM_StateTransition}. The Markov model then follows the law:

\begin{equation}
P(Q,X) \propto P(q_1) \prod_{t=2}^T P(q_t | q_{t-1}) \prod_{t=1}^T P(q_t | X_{p,t})
\end{equation}


where $\propto$ means ``proportional to", as the probabilities do not sum to 1. For $t=1,\ldots,T$, we assume that:

\begin{equation}
P(q_t=0 | q_t=0)=1-\tau_0 \quad P(q_t=1 | q_t=0)=\tau_0
\end{equation}
\begin{equation}
P(q_t=0 | q_t=1)=\tau_1 \quad P(q_t=1 | q_t=1)=1-\tau_1
\end{equation}

with $\tau_0,\tau_1 \in [0,1]$ the transition probabilities, and the convention that $q_0=0$ because all notes are inactive at the beginning of a recording. The transition probabilities $\tau$ correspond to the state transitions: on/on, on/off, off/on, off/off. Parameter $\tau_0$ (resp. $\tau_1$) is directly related to the prior duration of inactivity (resp. activity) of pitch $p$. Without observation, the length of an inactivity run (resp. activity run) would be geometric with parameter $\tau_0$ (resp. $\tau_1$) with average length $1/\tau_0$ (resp.  $1/\tau_0$). 

The observation probabilities are defined as follows, using a sigmoid curve with the PLCA pitch activity matrix $X_{p,t}$ as input,  

%

\begin{equation}\label{SigPlus}
P(q_t=0 | X_{p,t})\propto 1 / Z
\end{equation}
\begin{equation}\label{SigMoins}
P(q_t=1 | X_{p,t})\propto\exp\left[ e^\alpha (X_{p,t}-\beta) \right] / Z
\end{equation}


with $\alpha,\beta \in \mathbb R$, and $Z$ defined such as $\sum_{q_t} P(q_t | X_{p,t}) = Z$. The parameter of the model is denoted $\theta=(\tau,\alpha,\beta)$ which includes the specific value for all pitches. The HMM model is solved using classical forward-backward recursions for all $t=1,\ldots,T$, i.e. $P_\theta(q_{t}=s | X_{p,t})=\eta_s(t) \propto F_t(s)B_t(s)$. 

Note that the HMM definition combines both the spatial pitch dependence (the Markov model) with a PLCA generative model. As a result of this combination, the resulting model is defined up to a constant factor, but this is not a problem since we will exploit this model to compute posterior distribution. In contrast, in the initial \cite{Poliner2007}'s model, one should not that a similar model is suggested where the PLCA generative part is associated to so called ``virtual observation". We here preferred the fully generative formulation presented above, but both models are obviously totally equivalent.

\begin{figure}[htbp]
  \centering
  \includegraphics[width=\columnwidth]{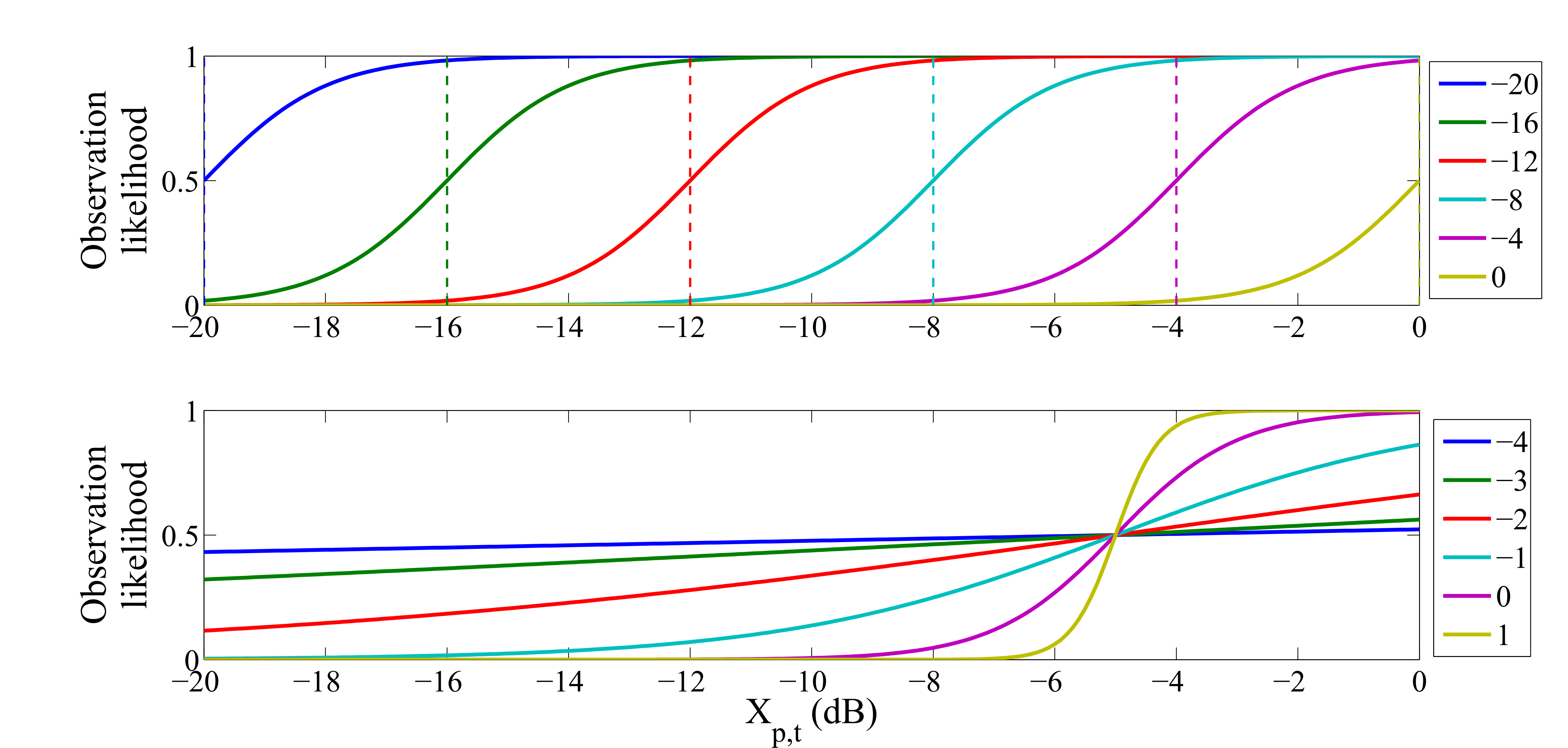}
  \caption{Effects of the parameters $\beta$ (top) and $\alpha$ (bottom) on the theoretical sigmoid given by Eq.~(\ref{SigMoins}). On top, a fixed value of 0 is set to $\alpha$, and on bottom, a fixed value of -5 is set to $\beta$.}
  \label{TheoreticalSigmoid}
\end{figure}

Using logarithmic values, the parameters $\{\alpha , \beta \}$, expressed in dB, are directly interpretable by physics. $\beta$ is an offset thresholding parameter, which allows separating signal from noise (or in other words, i.e. the higher its value, the more pitch candidates with low probability will be discarded.), while $\alpha$ is a contrast parameter, a value superior to 0 is used for a fast switch from noise to signal (i.e. low degree of tolerance from threshold), and a value inferior to 0 for a smoother switch. Figure \ref{TheoreticalSigmoid} shows a sigmoid curve with different values of $\beta$ and $\alpha$. This suggested parametrization $\{\alpha , \beta \}$ can therefore be seen as a generalization of the initial \cite{Poliner2007}'s model.

For this note segmentation strategy ST, we use the set of parameters $\{\alpha , \beta \}$ = $\{0 , \beta_\text{HT} \}$, as used in previous studies \cite{Poliner2007,Benetos2013}.


%
%

\subsubsection{OST: Optimized Soft Thresholding}\label{TwoStateHMMmodels}

\begin{figure}[htbp]
  \centering
  \includegraphics[width=\columnwidth]{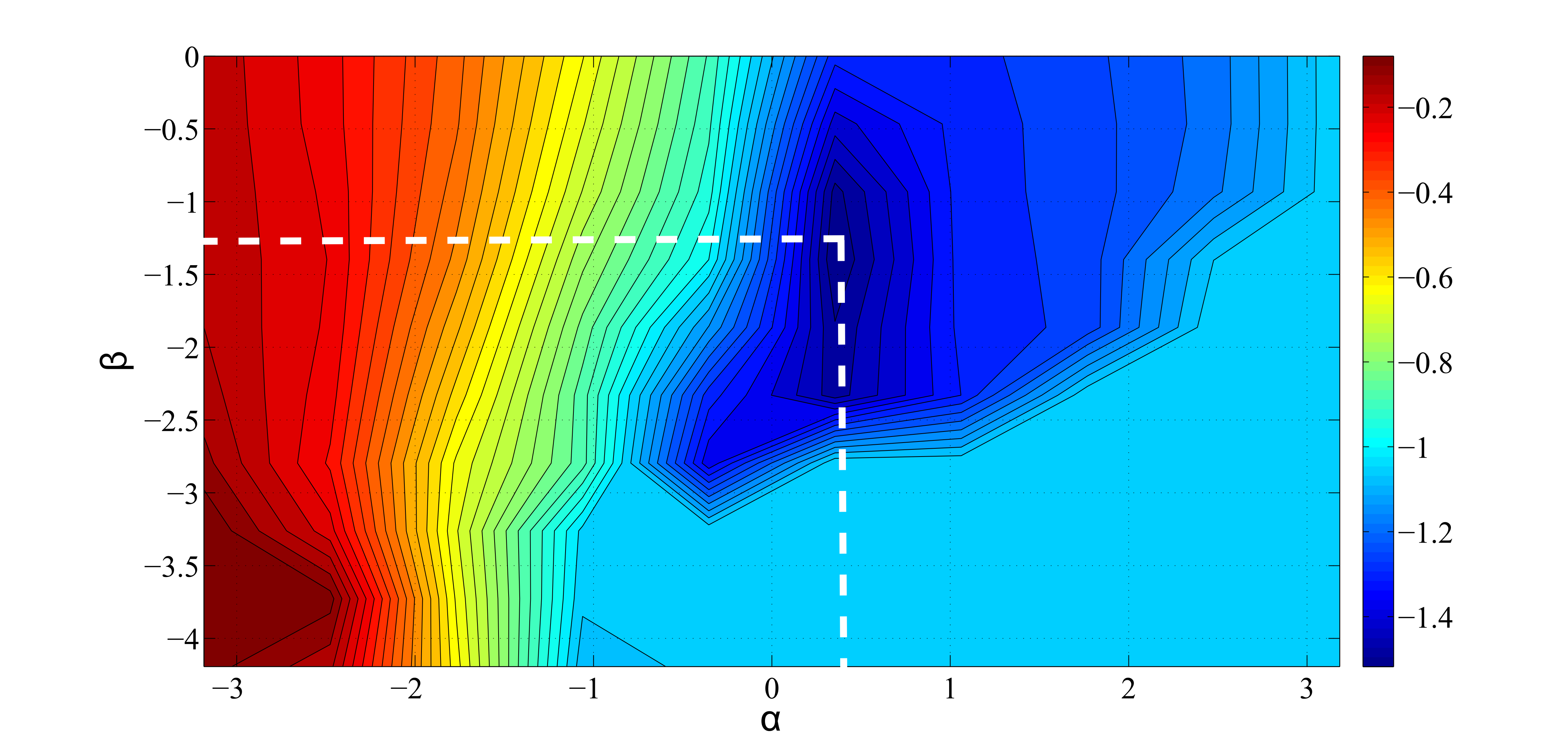}
  \caption{Example of a data-based optimization of the $\{\alpha , \beta \}$ parameters through the contour graph of the $\log_{10}(\text{LMSE})$ function, using the musical piece \texttt{MAPS\textunderscore MUS-alb\textunderscore esp2\textunderscore AkPnCGdD}. The dashed white lines point to the local minimum.}
  \label{ContourGraphOptimalAlphaBeta}
\end{figure}

The note segmentation strategy OST is based on the same HMM model as the ST strategy, although the parameters $\{\alpha , \beta \}$ are now optimized for each pitch. Given the ground truth of a musical sequence test, we use the Nelder-Mead optimizer of the R software to iteratively find the optimal $\{\alpha , \beta \}$ parameters that provide the best transcription performance measure. The Nelder-Mead method is a simplex-based multivariate optimizer known to be slow and imprecise but generally robust and suitable for irregular and difficult problems. For optimization, we use the Least Mean Square Error (LMSE) metric, as it allows to take into account the precise shape of activation profiles. Figure \ref{ContourGraphOptimalAlphaBeta} provides an example of this optimization through the contour graph of the $\log_{10}(\text{LMSE})$ function. However, classical AMT error metrics (see Sec. \ref{EvalMetricsAMT}) will be used as display variables for graphics as they allow direct interpretation and comparison in terms of transcription performance. 

In real world scenarios of AMT, the ground truth of a musical piece is never known by advance. A common strategy to estimate model or prior knowledge parameters is to train them on a learning dataset that is somewhat similar to the musical piece to be transcribed. This was done in this study for the $\{\alpha , \beta \}$ parameters, through a cross-validation procedure with the LMSE-optimization (see Sec. \ref{CrossValida}). 

\subsection{Evaluation procedure}

\subsubsection{Sound dataset}\label{SoundData}

To test and train the AMT systems, three different sound corpus are required: audio musical pieces of an instrument repertoire, the corresponding scores in the form of MIDI files, and a complete dataset of isolated notes for this instrument. Audio musical pieces and corresponding MIDI scores were extracted from the MAPS database \cite{Emiya2010}, belonging to the solo classical piano repertoire. The 56 musical pieces of the two pianos labelled AkPnCGdD and ENSTDkCl were used, and constituted our evaluation sound dataset called Baseline. The first piano model is the virtual instrument Akoustik Piano (concert grand D piano) developed by the software Native Instruments. The second one is the real upright piano model Yamaha Disklavier Mark III. Three other sound datasets of musical pieces have then been defined as follows:

\begin{itemize}
\item MP3 dataset. It corresponds to the same musical pieces of the dataset Baseline, but modified with the Strong MP3 Compression degradation from the degradation toolbox \cite{Mauch2013}. This degradation compresses the audio data to an MP3 file at a constant bit rate of 64 kbps using the Lame encoder ;
\item Smartphone dataset. It corresponds to the same musical pieces of the dataset Baseline, but modified with the Smartphone Recording degradation from the degradation toolbox \cite{Mauch2013}. This degradation simulates a user holding a phone in front of a speaker: 1. Apply Impulse Response, using the IR of a smartphone
microphone (``Google Nexus One"), 2. Dynamic Range Compression, to simulate the phone's auto-gain, 3. Clipping, 3 \% of samples, 4. Add Noise, adding medium pink noise ;
\item Vinyl dataset. It corresponds to the same musical pieces of the dataset Baseline, but modified with the Vinyl degradation from the degradation toolbox \cite{Mauch2013}. This degradation applies an Impulse Response, using a typical record player impulse response, adds Sound and record player crackle, a Wow Resample, imitating wow-and-flutter, with the wow-frequency set to 33 rpm (speed of Long Play records), and adds Noise and light pink noise.


\end{itemize}

For all datasets, isolated note samples were extracted from the RWC database (ref. 011, CD 1) \cite{Goto2003}.

\subsubsection{Cross-validation}\label{CrossValida}

During a cross-validation procedure, the model is fit to a training dataset, and predictive accuracy is assessed using a test dataset. Two cross-validation procedures were used for training the $\{\alpha , \beta \}$ parameters of the OST strategy, and testing separately the three thresholding strategies. The first one is the ``leave-one-out" cross-validation procedure, using only one musical piece for parameter training and testing all others. This process is iterated for each musical piece. The second one is a repeated random sub-sampling validation, also known as Monte Carlo cross-validation. At each iteration, the complete dataset of musical pieces is randomly split into training and test data accordingly to a given training/test ratio. The results are then averaged over the splits. The advantage of this method (over k-fold cross validation) is that the proportion of the training/test split is not dependent on the number of iterations (folds). A number of 20 iterations was used during our simulations. We also tested different training/test ratio, ranging from 10/90 \% to 60/40 \% in order to evaluate the influence of the training dataset on transcription performance.



\subsubsection{Evaluation metrics}\label{EvalMetricsAMT}

For assessing the performance of our proposed transcription system, frame-based evaluations are made by comparing the transcribed output and the MIDI ground-truth frame by frame using a 10 ms scale as in the MIREX multiple-$F_0$ estimation task \cite{Mirex2011}. We used the frame-based recall (\text{TPR}), precision (\text{PPV}), the F-measure (\text{FMeas}) and the overall accuracy (\text{Acc})

\begin{equation}\label{Metr1}
\text{TPR} = \frac{\sum_{t=1}^T  \text{TP}[t]}{\sum_{t=1}^T \text{TP}[t] + \text{FN}[t]} 
\end{equation}
\begin{equation}\label{Metr2}
\text{PPV} = \frac{\sum_{t=1}^T  \text{TP}[t]}{\sum_{t=1}^T  \text{TP}[t] + \text{FP}[t]}
\end{equation}
\begin{equation}\label{Metr4}
\text{FMeas} = \frac{2 . \text{PPV} . \text{TPR}}{ \text{PPV} + \text{TPR}} 
\end{equation}
\begin{equation}\label{Metr4}
\text{Acc} = \frac{\sum_{t=1}^T  \text{TP}[t]}{  \sum_{t=1}^T  \text{TP}[t] + \text{FP}[t]+ \text{FN}[t]} 
\end{equation}

where $T$ is the total number of time frames, and \text{TP}$[t]$, \text{TN}$[t]$, \text{FN}$[t]$ and \text{FP}$[t]$ are the numbers of true positive, true negative, false negative and false positive pitches at frame $t$. The recall is the ratio between the number of relevant and original items; the precision is the ratio between the number of relevant and detected items; and the F-measure is the harmonic mean between precision and recall. For all these evaluation metrics, a value of 1 represents a perfect match between the estimated transcription and the reference one.

\subsubsection{MPE algorithms on the benchmark}\label{AlgoBenchm}

In this study, we tested the four following MPE algorithms:

\begin{itemize}

\item Tolonen2000, this algorithm\footnote{We used the source code implemented in the MIR toolbox \cite{Lartillot2007}, called mirpitch(..., 'Tolonen').} \cite{Tolonen2000} is an efficient model for multipitch and periodicity analysis of complex audio signals. The model essentially divides the signal into two channels, below and above 1000 Hz, computes a ``generalized" autocorrelation of the low-channel signal and of the envelope of the high-channel signal, and sums the autocorrelation functions ;

\item Emiya2010, this algorithm\footnote{Source code courtesy of the primary author.} \cite{Emiya2010} models the spectral envelope of the overtones of each note with a smooth autoregressive model. For the background noise, a moving-average model is used and the combination of both tends to eliminate harmonic and sub-harmonic erroneous pitch estimations. This leads to a complete generative spectral model for simultaneous piano notes, which also explicitly includes the typical deviation from exact harmonicity in a piano overtone series. The pitch set which maximizes an approximate likelihood is selected from among a restricted number of possible pitch combinations as the one ;

\item HALCA, the Harmonic Adaptive Latent Component Analysis algorithm\footnote{Source codes are available at \url{http://www.benoit-fuentes.fr/publications.html}.} \cite{Fuentes2013} models each note in a constant-Q transform as a weighted sum of fixed narrowband harmonic spectra, spectrally convolved with some impulse that defines the pitch. All parameters are estimated by means of the EM algorithm, in the PLCA framework. This algorithm was evaluated by MIREX and obtained the $2^{nd}$ best score in the Multiple Fundamental Frequency Estimation $\&$ Tracking task, 2009-2012 \cite{Mirex2011} ;

\item Benetos2013, this PLCA-based AMT system\footnote{Source codes are available at \url{https://code.soundsoftware.ac.uk/projects/amt_mssiplca_fast}.} \cite{Benetos2013d} uses pre-fixed templates defined with real note samples, without updating them in the maximization step of the EM algorithm. It has been ranked first in the MIREX transcription tasks \cite{Mirex2011}. 

\end{itemize}


\subsection{Setting the HT threshold value}

We need to define the threshold value $\beta_\text{HT}$ used in the note segmentation strategies HT and ST. Although most studies in AMT literature \cite{Mysore2009,Dessein2010,Grindlay2011} use this note segmentation strategy, threshold values are barely reported and procedures to define them have not yet been standardize. Most of the time, one threshold value is computed across each evaluation dataset, which is dependent on various parameters of the experimental set-up, such as the used evaluation metric, input time-frequency representation, normalization of input waveform. In this paper, we will use a similar empirical dataset-based approach to define the HT threshold value. ROC curves (True Positives against False Positives) are computed over the threshold range [0 ; -5] dB so as to choose the value that maximizes True Positive and minimizes False Positives, i.e. that increases transcription performance at best over each dataset.

%
%
%

\section{Results and Discussion}

\begin{figure}[htbp]
  \centering
  \includegraphics[width=0.95\columnwidth]{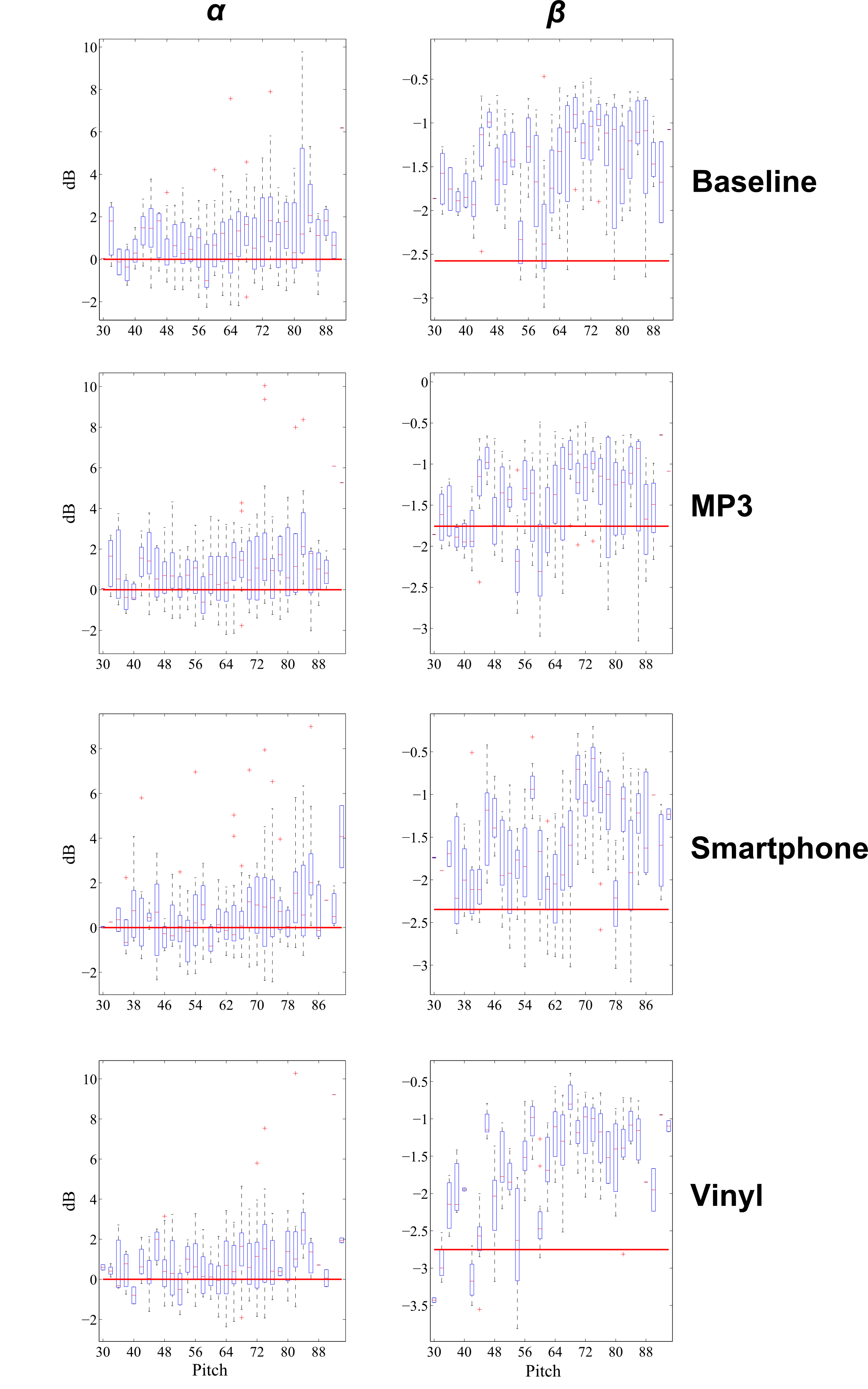}
  \caption{Boxplots of the optimal $\{\alpha , \beta \}$ values obtained for each pitch, and for each evaluation dataset. The horizontal red lines in each boxplot represents the parameter values used in the ST and HT thresholding strategies.}
  \label{Boxplot_OptimalAlphaPerPitch}
\end{figure}

\begin{figure}[htbp]
  \centering
  \includegraphics[width=\columnwidth]{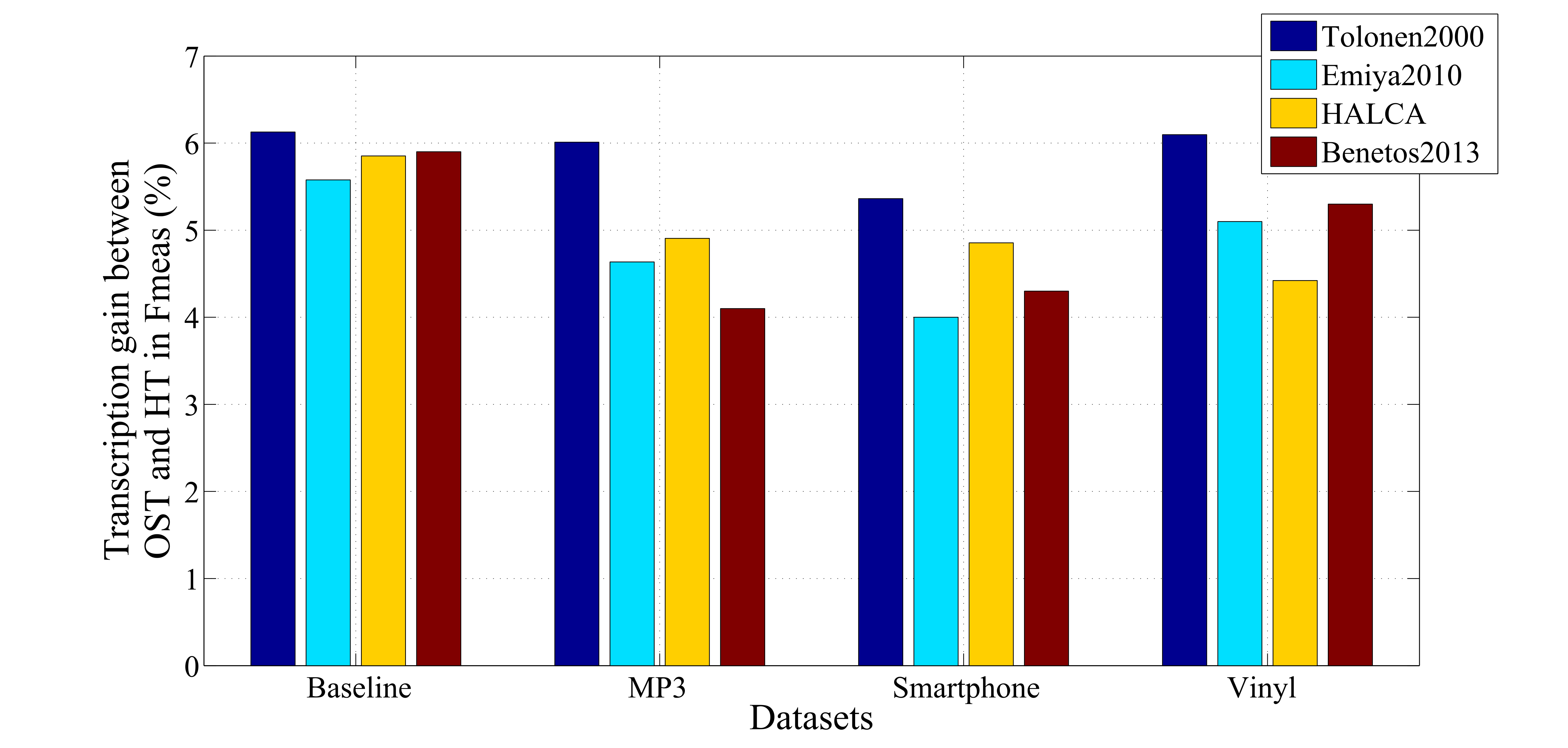}
  \caption{Difference between the F-measures obtained with the OST and HT note segmentation methods, using different AMT systems.}
  \label{SensiStudyAMTsystems}
\end{figure}

All following results on transcription performance have been obtained using the Benetos2013 AMT system, except for figure \ref{SensiStudyAMTsystems} where all AMT systems are comparatively evaluated. Figure \ref{Boxplot_OptimalAlphaPerPitch} represents the boxplots of the optimal $\{\alpha , \beta \}$ values obtained for each pitch. The ``leave-one-out" cross-validation procedure has been applied to the different datasets, from top to bottom. For each dataset, we can see that the data-based pitch-wise optimization leads to $\beta$ values drastically different from the threshold value $\beta_\text{HT}$ used in the ST and HT thresholding strategies (represented by the horizontal red lines). Differences range from 0.5 to 2 dB, that have a significant impact for note segmentation. Slighter differences are observed in values of $\alpha$, although slightly positive values of $\alpha$ (around + 1 dB) tend to contribute to reduce the LMSE metric used in optimization. Also, note that optimal $\beta_\text{HT}$ values are also dependent on the datasets, varying from -1.8 to -2.8 dB. 

Now, let's see how this optimization of $\{\alpha , \beta \}$ in the method OST impacts real transcription performance. Table \ref{TestPerformance_LOU} shows transcription results obtained with the ``leave-one-out" cross-validation procedure, applied to the different thresholding strategies. In comparison to the methods HT and ST, significant gains in transcription performance are brought by the proposed method OST. These gains are the highest for the baseline dataset $D_1$, in the order of magnitude of 5 to 8 \% for the two metrics \text{Acc} and \text{FMeas}. They remain systematically positive for the other datasets, with a minimum gain of 4 \% whatever the dataset, error metric and compared thresholding strategy. Altogether, these gains are very significant in regards to common gains in transcription performance reported in literature, and demonstrate the validity of our proposed method.

\begin{table}[h]
\centering
\resizebox{0.95\columnwidth}{!}{

\begin{tabular}{|c|c|c|c|}
\hline
Datasets                    & \begin{tabular}[c]{@{}c@{}}Note segmentation\\ strategies\end{tabular} & Acc (\%)      & Fmeas (\%)    \\ \hline
\multirow{3}{*}{Baseline}   & HT                                                                     & 54.9          & 53.3          \\ \cline{2-4} 
                            & ST                                                                     & 57.6          & 55.3          \\ \cline{2-4} 
                            & \textbf{OST}                                                           & \textbf{62.3} & \textbf{59.2} \\ \hline
\multirow{3}{*}{MP3}        & HT                                                                     & 51.9          & 52.6          \\ \cline{2-4} 
                            & ST                                                                     & 52.2          & 50.1          \\ \cline{2-4} 
                            & \textbf{OST}                                                           & \textbf{55.6} & \textbf{56.7} \\ \hline
\multirow{3}{*}{Smartphone} & HT                                                                     & 52.2          & 51.9          \\ \cline{2-4} 
                            & ST                                                                     & 53.1          & 51.3          \\ \cline{2-4} 
                            & \textbf{OST}                                                           & \textbf{58.4} & \textbf{56.5} \\ \hline
\multirow{3}{*}{Vinyl}      & HT                                                                     & 50.8          & 48.8          \\ \cline{2-4} 
                            & ST                                                                     & 51.1          & 49.2          \\ \cline{2-4} 
                            & \textbf{OST}                                                           & \textbf{57.8} & \textbf{54.1} \\ \hline
\end{tabular}

}
\caption{Averages of error metrics \text{FMeas} and \text{Acc} obtained with the different thresholding strategies, i.e. ST,  OST and HT, using a leave-one-out cross-validation procedure.}\label{TestPerformance_LOU}
\end{table}

%
%
%
%

\begin{figure}[htbp]
  \centering
  \includegraphics[width=\columnwidth]{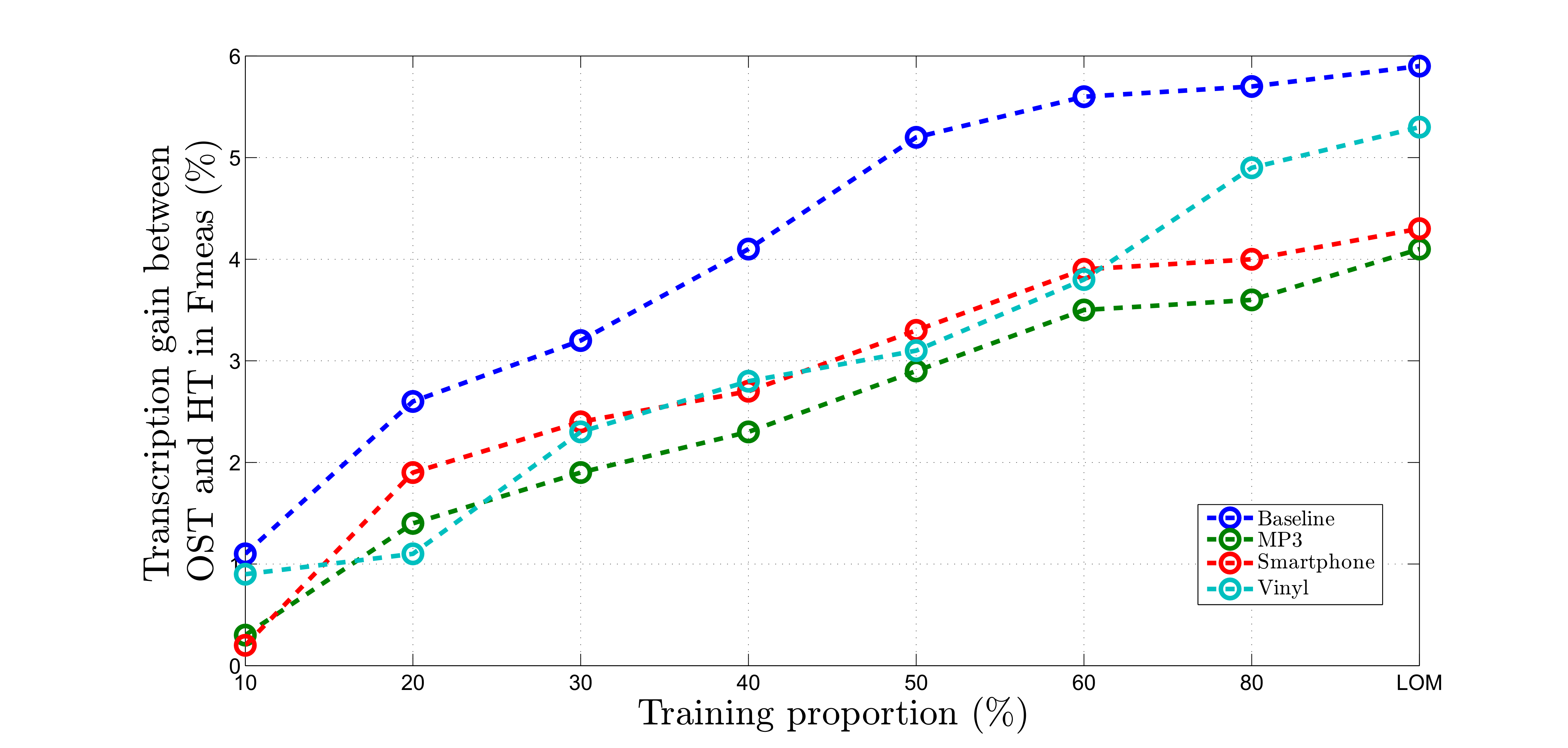}
  \caption{Difference between the F-measures obtained with the OST and HT note segmentation methods, using 20 iterations of the repeated random sub-sampling validation method with training/test ratio ranging from \textbf{10}/90 \% to \textbf{60}/40 \%, plus the ``leave-one-out" (labelled LOM) ratio.}
  \label{EvolutionPerfoWithTrainingSize}
\end{figure}

In Figure \ref{EvolutionPerfoWithTrainingSize}, we evaluated the dependency of transcription performance on the training dataset size, through a Monte Carlo cross-validation procedure with different training/test ratios, ranging from 10 to 60 \% of the complete dataset of musical pieces, plus the ``leave-one-out" (labelled LOM) ratio. This figure shows that increasing the size of the training set directly induces average transcription gains from 0.5 to 6 \% of the metric \text{FMeas} with the OST method, in comparison to the HT method. We note that once the curves reach the \textbf{60}/40 \% training/test ratio, all systems find a quick convergence to the gain ceiling achieved with the LOM ratio.

Eventually, we studied the dependency of OST transcription performance on the AMT system used, in comparison to the method HT. Figure \ref{SensiStudyAMTsystems} shows the differences between the FMeas obtained with the methods OST and HT. We can observe that these differences are relatively small, i.e. inferior to 2 \%. This demonstrates that the proposed OST method improves transcription performance in a rather universal way, as independent from the characteristics of activation matrices as long as AMT system specific training datasets are used. Only AMT system Tolonen2000 shows higher transcription gains (especially for the datasets $D_3$ and $D_4$) brought by the OST method as this system outputs the worst activation matrices.

\section{Conclusion}

In this study, an original method for the task of note segmentation was presented. This task is a crucial processing step in most systems of automatic music transcription. The presented method is based on a two-state pitch-wise Hidden Markov Model method, augmented with two sigmoid parameters on contrast and slope smoothing that are trained with a learning dataset. This rather simple method has brought significant results in transcription performance on music datasets with different characteristics. It can also be used as a universal post-processing block after any pitch-wise activation matrix, showing great promise for future use.


%
%
%
%

\end{document}